\documentclass[journal=jacsat,manuscript=article]{achemso}
\usepackage{lineno}
\usepackage{amsmath}
\usepackage{gensymb}
\DeclareMathOperator{\Tr}{Tr}
\usepackage{xcolor}
\newcommand{\update}[1]{#1}

\SectionNumbersOn

\title{Efficient inverse design of large-area metasurfaces for incoherent light}

\author{Rapha{\"e}l Pestourie}
\affiliation{Mathematics Department, Massachusetts Institute of Technology, Cambridge, MA 02139, United States}
\email{rpestour@mit.edu}
\author{Wenjie Yao}
\affiliation{Electrical Engineering and Computer Science Department, Massachusetts Institute of Technology, Cambridge, MA 02139, United States}

\author{Boubacar Kant{\'e}}
\affiliation{Department of Electrical Engineering and Computer Sciences, University of California, Berkeley, CA 94720, United States}
\author{Steven G. Johnson}
\affiliation{Mathematics Department, Massachusetts Institute of Technology, Cambridge, MA 02139, United States}
\email{stevenj@math.mit.edu}

\begin{document}

\begin{abstract}
% new frontier or critical application
Incoherent light is ubiquitous, yet designing optical devices that can handle its random nature is very challenging, since directly averaging over many incoherent incident beams can require a huge number of scattering calculations.
We show how to instead solve this problem with a reciprocity technique which leads to three orders of magnitude speedup: one Maxwell solve (using any numerical technique) instead of thousands.  This improvement enables us to perform efficient inverse design, large-scale optimization of the metasurface for applications such as light collimators and concentrators.  We show the impact of the angular distribution of incident light on the resulting performance, and show especially promising designs for the case of ``annular'' beams distributed only over nonzero angles.
% We present a technique that can very efficiently compute the average effect of random light with several of magnitude speedup compared to intuitively averaging forward simulations. We showcase the power for this framework with optimize metasurfaces that act as a concentrator and a collimator for incoherent light of different distributions.
\end{abstract}

\section{Keywords}
Inverse design; Incoherent light; Collimator; Concentrator; Metasurface

%%%%%%%%%%%%%%%%%%%%%%%%%%  body  %%%%%%%%%%%%%%%%%%%%%%%%%%
\section{Introduction}

\begin{figure}
    \centering
    \includegraphics[width=\textwidth]{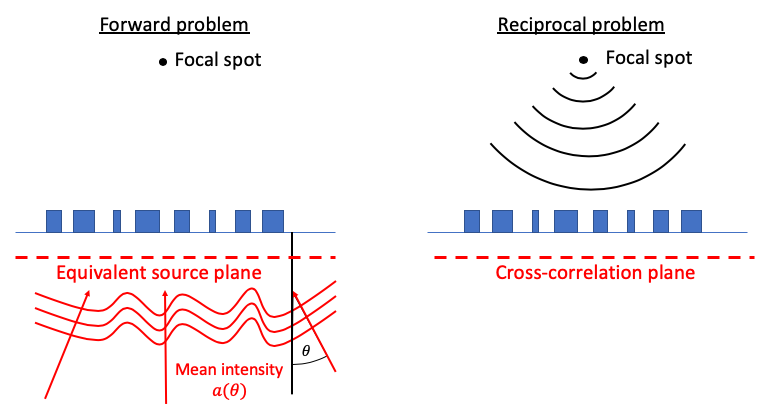}
    \caption{(left) Concentrator forward problem: incoherent light is incident on a metasurface and the mean intensity of the scattered field is measured at the focal spot. The incoherent light is characterized by a distribution of the mean intensity $a(\theta)$ of a mix of planewaves with incident angles in a predefined interval $\theta\in I$. The incident light is transformed into an equivalent source in the forward simulation using the principle of equivalence. The average intensity at the focal spot is computed with a numerical integral on the interval $I$, which requires many Maxwell solves (1000s) with different angles of incidence to be accurate. (right) Concentrator reciprocal problem: a point source is placed at the focal spot and the scattered field through the metasurface is measured and convoluted with the spectral-cross correlation. This method computes the average intensity at the focal spot with a single Maxwell solve. For a collimator application, the source would be the mode of the monitor.}
    \label{fig:diagram}
\end{figure}

We report a fast reciprocity-based method to both model and inverse-design \emph{averaged} focusing or collimation properties of metasurfaces (large-area subwavelength-patterned aperiodic structures for free-space optics) for \emph{incoherent} light with any prescribed statistics, and show that it is orders of magnitude faster than direct angle averaging.
A naive approach would compute the average effect of incoherent light with a numerical integration (quadrature) scheme over incident angles for simulations with planewave sources and a given intensity distribution. By instead introducing a ``reciprocal'' problem based on a general framework for incoherent-light optimization derived in recent work~\cite{yao2021trace}, we reduce the computation to a single Maxwell solve (in 2D, or three solves in 3D) with a prescribed source followed by a convolution of the resulting electromagnetic (EM) field with a cross-correlation function of the incoherent light (Fig.~\ref{fig:diagram}, Sec.~\ref{sec:math}). We apply this approach to metasurface designs for collimator (single output angle) and concentrator (single focal spot) applications, considering incident light distributed uniformly in either a cone (a single range of angles around $0\degree$) or an annulus (ranges of nonzero angles excluding $0\degree$).
Compared to a baseline single-angle lens or collimator, optimization yields
especially dramatic improvements ($12\times$ or $2.6\times$, respectively) in the more unusual case of annular incoherent beams (Fig.~\ref{fig:designs}).  We also find that the angular dependence of optimizing \emph{average} performance can be very different from optimizing the \emph{worst-angle} performance as in some earlier concentrator work~\cite{lin2021computational, roques2022toward}, except for very large angular ranges ($> 40\degree$).   Although previous designs sampled only 6--20 angles\cite{lin2021computational, roques2022toward, scranton2014single}, we find that simulations for $\sim 10^3$ angles would be required to obtain an accurate brute-force average over $\pm 20\degree$, corresponding to a three-order-of-magnitude speedup for our reciprocal approach. (Even larger speedups will occur in 3D, where two incident angles must be considered in general.) We believe that this technique will be crucial to developing metasurfaces and photomasks for incoherent-light applications ranging from lithography~\cite{scranton2014single} to bioluminescence sensing (as discussed in Sec.~\ref{sec:ccl}).

% Claims: 
% \begin{itemize}
% \item solving for average effect of incoherent light only requires a reciprocal simulation: point source for concentration and planewave for collimator, and a convolution with the cros-spectral desnity of the incoherent light
% \item example design of a concentrator that can improve over a lens design.
% \item example design of a collimator can improve collimation by up to 500\% compared to a periodic metasurface baseline, in the case of incoherent annular light.
% \end{itemize}

\section{Incoherent light}\label{sec:math}

We first consider the ``forward'' problem where incoherent light is incident on a metasurface.  For simplicity, we will illustrate our approach in 2D, where electromagnetic waves can be defined by a scalar field $u$ (either the out-of-plane electric or magnetic field component, depending on the polarizatin).   Incident incoherent light $u_\mathrm{inc}$ can then be described using an angular spectrum representation~\cite{wolf2007introduction, yablonovitch1982statistical} as a superposition of planewaves with random amplitudes $a(\theta)$ for all incident angles $\theta$ in the interval of allowed angles of the incoherent light~$I$
\begin{equation}\label{eq:expansion}
    u_\mathrm{inc} = \int_{\theta\in I} d\theta a(\theta) \exp(ik(\theta)r) .
\end{equation}
The amplitude is assumed to have zero mean (i.e. $a$ and $-a$ are equally likely).
The statistics of the incoherent light are defined by the autocorrelation function of the (complex) angle-dependent amplitude $a$
\begin{equation}\label{eq:autocorrelation}
    \mathrm{Corr}(a(\theta), a(\theta^\prime))=\langle a^*(\theta)a(\theta^\prime) \rangle
\end{equation}
where $^*$ denotes complex conjugation, and $\langle \cdots \rangle$ denotes the ensemble average with respect to the joint probability density of the random amplitude pair~$(a(\theta), a(\theta^\prime))$.   We will show below that this further simplifies in the common case of broad illumination with translation-invariant spatial correlations, in which case the amplitudes at different angles are \emph{uncorrelated} and one only needs the mean-square amplitude $\langle |a(\theta)|^2 \rangle$ at each angle of incidence.

By the principle of equivalence~\cite{oskooi2013electromagnetic}, any incident wave can be equivalently produced by \update{a current source} distribution $b$ defined an enclosing surface, such as the source plane depicted Fig.~\ref{fig:diagram}(left).
Random planewaves can then be tranformed to random equivalent currents, whose spatial cross-correlation~\cite{wolf2007introduction} %$\langle bb^\dag \rangle$ 
is
\begin{equation}\label{eq:crosscorrelation}
    W(x_1, x_2) = \int_{(\theta, \theta^\prime) \in I} d\theta d\theta^\prime \langle a(\theta) a^*(\theta^\prime) \rangle \exp(i k_x(\theta^\prime) x_2) \exp(-i k_x(\theta) x_1)
\end{equation}
with $k_x(\theta)=n\frac{2\pi}{\lambda}\sin{\theta}$, where $n$ is the refractive index of the medium. 

For spatially broad incoherent illumination, the spatial correlation function is commonly approximated as being translation invariant, i.e. $W(x_1, x_2) = W(x_1 - x_2)$ is only a function of $x_1 - x_2$.   Equivalently, in the Fourier domain, translation-invariant correlations are diagonalized: different Fourier components, corresponding to planewaves at different angles, are uncorrelated.   Consequently, in this regime we have $\mathrm{Corr}(a(\theta), a(\theta^\prime))=\frac{1}{2c} \delta(\theta-\theta^\prime) \langle |a(\theta)|^2 \rangle$ (where $\delta$ is the Dirac delta function): the incident light can be viewed as a distribution of \emph{uncorrelated} random planewaves at different angles with some mean-square intensities.  This is the regime that we will consider for the specific computational examples in this paper, but the reciprocity-based computational technique (below) is also applicable to the more general situation of arbitrarily correlated incoherent light.

% in incoherent light section
A simple example of incoherent illumination is light distributed uniformly in a cone $I$ of angles: $\theta \in [-c, c] = I$, with equal mean-square amplitude at every angle.
Mathematically, $\mathrm{Corr}(a(\theta), a(\theta^\prime))=\frac{1}{2c} \delta(\theta-\theta^\prime)$,
or $\langle |a(\theta)|^2 \rangle = 1/2c$ (so that the total mean power is fixed as we vary $c$).  This corresponds to a spatial correlation
\begin{equation}\label{eq:W}
    W(x_1, x_2) =
    \frac{1}{2c}\int_{-c}^c d\theta \exp(i k_x(\theta) (x_2-x_1))
    =\mathrm{sinc}\left((x_2-x_1)\frac{2\pi}{\lambda}\sin{c}\right) .
\end{equation}
As the angular interval $c$ approaches $90\degree$, the spatial correlations $W$ become narrower and narrower, tending to $\frac{\lambda}{2}$. For example, in Ref.~\citenum{morrill2016measuring}, they experimentally measured a coherence length of $330$~nm for a wavelength of $\lambda=500$~nm with a range of angle $I=[-30\degree, 30\degree]$.

\section{Method}
We want to compute the averaged intensity of the scattered field at the focal spot in the case of a concentrator application (as in Fig.~\ref{fig:diagram}(left)), and the power in the collimator mode---a planewave with a desired transmitted angle---in the case of a collimator application. 
A naive way to compute the average, for the simple case of angle-uncorrelated planewaves, would be first to simulate the scattered field for each planewave angle, and then to average the result over angles weighted by $\langle |a(\theta)|^2 \rangle$, or more generally for some $a$ correlation function (math details in Sec.~\ref{sec:mathdetails}). However, this approach is computationally expensive, because we show below that the integrand is a highly oscillatory function of $\theta$ for large-area metasurfaces, and hence thousands of simulations may be required to compute an accurate average (Sec.~\ref{sec:results}).

In this work, we present an efficient technique to compute the average effect with a single ``reciprocal'' simulation. The source of the reciprocal simulation depends on the design objective: a point source for a concentrator application (as shown in Fig.~\ref{fig:diagram}(right)) or a plane wave at the desired angle for a collimator application. The resulting reciprocal fields are then convolved with the source cross-correlation function on the same plane as the equivalent source plane (math details in Sec.~\ref{sec:mathdetails}). The speedup benefit from the reciprocal simulation is independent of the choice of numerical technique to solve Maxwell's equations for the reciprocal scattering problem.

In this article, we limit our applications to two-dimensional simulations, where the interface of the metasurface is along the x-direction, but the same approach would apply to three-dimensional problems as discussed in Sec.~\ref{sec:ccl}.   We consider the frequency-domain problem of solving/designing the effect of incoherent light at a single wavelength, but of course this calculation could be repeated at multiple wavelengths to model broadband problems. 

\subsection{Reciprocal approach}\label{sec:mathdetails}

For a given wavelength, we denote the solutions to Maxwell's equations by $u = A^{-1}b$ where $A$ is the Maxwell's equations operator for the metasurface and $b$ is a current-source term.  (Alternatively can be convenient to think of $A$ as a matrix coming from a huge finite-difference or finite-element discretization of Maxwell's equations.)
We let $b_0(\theta)$ denote the equivalent current source for an unit-power incident planewave at an angle $\theta$, via the principle of equivalence~\cite{oskooi2013electromagnetic}, yielding solutions $u_0(\theta) = A^{-1} b_0(\theta)$. As in \eqref{eq:expansion} above, incoherent light is modeled as a source term $b$ given by a random superposition of $b_0(\theta)$ sources weighted according to some $a$ correlation function.

The figure of merit (FOM) of both a concentrator and a collimator share a common form of a squared inner product
$|w^\dag u|^2$,
where $u=\int_{\theta\in I} d\theta  a(\theta)u_0(\theta)$ is the scattered incoherent light, $w$ is the function that characterizes the application, and $w^\dag u$ is an inner product (integral). For concentrator (lens-like) application, $w$ is a Delta function at the position of the focal spot, so that $w^\dag u$ is simply the field $u$ evaluated at the focal spot as in Fig.~\ref{fig:diagram}. For a collimator application, $w$ is the desired output mode profile (e.g. a planewave at a desired angle), so that $w^\dag u$ is the amplitude of the desired output mode and $|w^\dag u|^2$ is the corresponding power.

For incoherent light, we then want the \emph{average} FOM:
\begin{align}\label{eq:forward}
\begin{split}
\langle |w^\dag u|^2 \rangle &= \left\langle  \left| w^\dag \left(\int_{\theta\in I} d\theta a(\theta)u(\theta)\right)\right|^2 \right\rangle \\
&= \left\langle  w^\dag \left(\int_{\theta\in I} d\theta a(\theta)u(\theta)\right) \left(\int_{\theta\in I} d\theta' a^*(\theta')u(\theta')^\dag\right)w\right\rangle\\
&= \left\langle \int_{(\theta, \theta^\prime) \in I} d\theta d\theta^\prime  a(\theta)a^*(\theta^\prime) w^\dag u(\theta)u(\theta^\prime)^\dag w\right\rangle \\
&= \int_{(\theta, \theta^\prime) \in I} d\theta d\theta^\prime  \langle a(\theta)a^*(\theta^\prime)\rangle w^\dag u(\theta)u(\theta^\prime)^\dag w \, .
\end{split}
\end{align}
For the simplified case of uncorrelated angles, this becomes
\begin{align}\label{eq:forwarduncorr}
\int_{\theta \in I} d\theta  \langle |a(\theta)|^2 \rangle  w^\dag u(\theta)u(\theta)^\dag w \, .
\end{align}
To evaluate \eqref{eq:forward} or \eqref{eq:forwarduncorr} by numerical integration, a forward simulation of the entire metasurface would be needed to compute $u(\theta)$ (and hence $u(\theta^\prime)$) for each quadrature point~$\theta$. As we show below, this may require so many Maxwell solves as to be prohibitive even for simple incoherent light probability distributions.

Similarly to Ref.~\citenum{yao2021trace}, we rewrite \eqref{eq:forward} using the cyclic property of the trace function and the linearity of the ensemble average:
\begin{align}\label{eq:reciprocal}
\begin{split}
    \langle |w^\dag u|^2 \rangle &= \langle  b^\dag A^{-\dag}w w^\dag A^{-1}b \rangle = \Tr\left(\langle  b^\dag A^{-\dag}w w^\dag A^{-1}b \rangle\right)  \\
    &= \Tr\left(\langle A^{-\dag}w w^\dag A^{-1}bb^\dag \rangle\right) = \Tr\left( A^{-\dag}w w^\dag A^{-1}\langle bb^\dag \rangle\right)\\
    &= \Tr\left(w^\dag A^{-1}\langle bb^\dag \rangle A^{-\dag}w \right) = \Tr\left(v^\dag\langle bb^\dag \rangle v \right) = v^\dag\langle bb^\dag \rangle v = v^\dag W v\\
    &=\int dx_1 dx_2 v(x_1)^\dag W(x_1, x_2) v(x_2)\\ 
\end{split}
\end{align}
where $\langle bb^\dag \rangle$ is the current--current correlation operator (the correlation ``matrix'').
In contrast to \eqref{eq:forward}, this new formulation requires only a \emph{single} solve of Maxwell's equations for a ``reciprocal'' simulation $v = A^{-\dag}w$, where the \emph{output} $w$ is treated as a ``source.'' \update{To optimize the figure of merit with respect to the many parameters $p$, we apply gradient-based optimization~\cite{pestourie2020assume, li2022empowering}. The gradient of the figure of merit is $2\Re\left(v^\dag W\nabla_p v\right)$, where $\nabla_p v$ is the Jacobian of the reciprocal fields on the cross-correlation plane.
In the case of our local periodic approximation (LPA, see below) $\nabla_p v$ ends up being diagonal (each unit cell only affects the near field directly below it), so we can compute it explicitly. More generally, one could apply an adjoint method~\cite{li2022empowering, pestourie2020assume} in which the Maxwell solve $A^{-\dag}$ in $\nabla_p v=- A^{-\dag}   \nabla_p A^{\dag}  v$ is applied to the left on a ``source'' term $v^{\dag} W$. The Maxwell operator $A$ is a composition~\cite{pestourie2018inverse, pestourie2020assume} of a Green’s function, which does not depend on the geometry $p$, and the local metasurface transmission for each unit cell as a function of the nanopillar diameter, which is computed from a surrogate polynomial fit~\cite{pestourie2018inverse, pestourie2020assume}. Hence the derivative $\nabla_p A^{\dag}$ merely involves the derivatives of these polynomials.} 

In the case of a concentrator application, $w$ is a point source situated at the focal spot. In the case of a collimator application, $w$ is a planewave source incident from the \emph{output} direction/mode. 
The correlation $W$ is a known function (often computed analytically) for any given incoherent-light distribution as described above.

% keep FFT here
In contrast to Ref.~\citenum{yao2021trace}, which primarily involved spatially uncorrelated random sources, the random currents here are correlated in space. In our examples below, we assume that the amplitudes are uncorrelated for different angles (corresponding to spatially broad and translation-invariant illumination) as described above.
In this case, $W(x_1, x_2) = W(x_2-x_1)$) so computing the average then boils down to a \emph{convolution} of $W$ with the reciprocal fields evaluated on the equivalent source plane as in \eqref{eq:reciprocal}, which can be efficiently computed numerically using fast Fourier transforms (FFTs)~\cite{frigo2005design}.

\subsection{Fast metasurface solver}
The primary benefit of the reciprocal-source technique we introduced above is a vast reduction in the number of Maxwell solves needed to compute the average effect of incoherent light. This benefit is independent of the computational technique used to solve Maxwell's equations. 

In this work, we perform our metasurface simulation and optimization using an approximate solver based on LPA as described in previous work~\cite{pestourie2020assume}. But of course, many other computational techniques for metasurfaces have been proposed~\cite{hsu2017local, LinJo19,skarda2022low, hughes2021perspective, phan2019high}, which could also be used with the same reciprocal approach.

The collimator application below and the forward simulations use planewave sources similar to Ref.~\citenum{pestourie2020assume}. For the reciprocal simulation of the concentrator application, the source is a point source, and a paraxial approximation~\cite{born2019principles} was found to be accurate to a few percent: the local fields adjacent to the metasurface can be computed using subdomain simulations (for each unit cell in LPA) with normal-incidence light whose phase is modulated according to the distance between the point source and the unit cell. 

%The local field simulations are normalized to unit incident power. By the principle of equivalence [ref] this means that the source and mode of the monitor have amplitudes proportional to $\sqrt{n\cos{\theta}}$ where $n$ is the refractive index of the medium and $\theta$ is the angle of propagation.

\begin{figure}
    \centering
    \includegraphics[width=\textwidth]{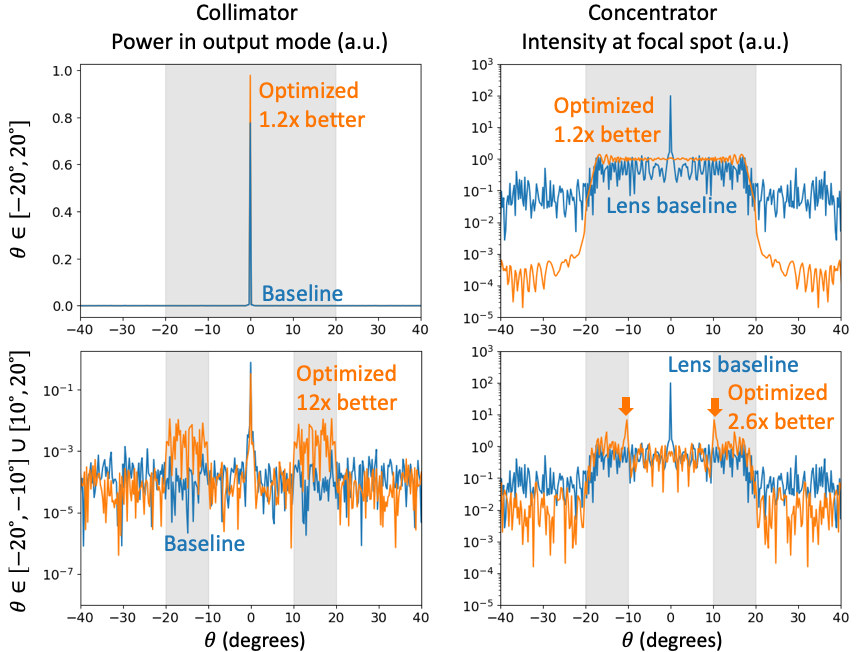}
    \caption{(top row) Example designs of a collimator and a concentrator for a uniform angle distribution on the interval $\theta \in [-20\degree , 20\degree]$ (shaded area). (left) power in the output mode of the collimator (in arbitrary units) in function of the angle of incidence, the optimized structure has a $1.18\times$ better figure of merit, averaged on the interval of the shaded region. (right) semilog plot of the intensity at the focal spot in function of the angle of incidence, the optimized structure performs consistently across angles between $-15\degree$ and $15 \degree$ and has a $1.2\times$ better figure of merit, averaged on the interval of the shaded region. (bottom row) Example designs of a collimator and a concentrator for a uniform angle distribution on a union of disjoint intervals $\theta \in [-20\degree , -10\degree]\cup [10\degree , 20\degree]$ (shaded areas). (left) The power output mode is significantly improved consistently across the two intervals of interest, leading to a $12\times$ improvement of the average power in the output mode of the collimator. (right) The optimized concentrator, significantly increases the the intensity for $-10\degree$ and $10\degree$, which are the angles from both intervals that are closest to the normal incidence, leading to a $2.6\times$ improvement.}
    \label{fig:designs}
\end{figure}

\section{Results}\label{sec:results}
\subsection{Designs}

We show example designs of collimators and concentrators. As in Ref.~\citenum{bayati2022inverse}, the metasurface consists of pillars made of silicon nitride ($\varepsilon=4$) on top of a silica substrate ($\varepsilon=2$). The wavelength of incident light is $\lambda=633$~nm and the subwavelength period of the unit cell is 316~nm. The minimum feature size of the pillars is 100~nm which results in a parameter range of [100, 216]~nm. The metasurface consists of $10^3$ unit cells. For the concentrator design with a numerical aperture of $0.3$, we compare to the baseline of a lens designed to focus normal-incident light only. For the collimator, we compare to the baseline of a random metasurface.

In Fig.~\ref{fig:designs}(top row), we show example designs of a collimator and a concentrator for a uniform angle distribution on the interval $\theta \in [-20\degree , 20\degree]$ (shaded area) using the cross-correlation from \eqref{eq:W}. In Fig.~\ref{fig:designs}(left) we show the power in the output mode of the collimator (in arbitrary units) in function of the angle of incidence, the optimized structure performs about 25\% better for normal incidence resulting in a $1.18\times$ improvement of the averaged figure of merit. In Fig.~\ref{fig:designs}(right), we show the semilog plot of the intensity at the focal spot in function of the angle of incidence, the optimized structure performs consistently across angles between $-15\degree$ and $15 \degree$, resulting in a $1.2\times$ improvement of the averaged figure of merit.  \update{Note that the optimized structure is on average two times better than the lens in the 1---18$\degree$ range away from the central peak. This is achieved, in part, by trading off performance at unneeded angles $>20\degree$.}

In Fig.~\ref{fig:designs}(bottom row), we show example designs of a collimator and a concentrator for a uniform angle distribution on a union of disjoint intervals $\theta \in [-20\degree , -10\degree]\cup [10\degree , 20\degree]$ (shaded areas) in which case the cross-correlation function becomes
\begin{equation}
    W(x_1, x_2) = \frac{1}{(\theta_2-\theta_1)} \left(\theta_2 \mathrm{sinc}\left((x_2-x_1)\frac{2\pi}{\lambda}\sin{\theta_2}\right) - \theta_1 \mathrm{sinc}\left((x_2-x_1)\frac{2\pi}{\lambda}\sin{\theta_1}\right) \right)
\end{equation} where $\theta_2 = 20\degree$ and $\theta_1=10\degree$ In Fig.~\ref{fig:designs}(left), we show the power output mode is significantly improved consistently across the two intervals of interest, leading to a $12\times$ improvement of the average power in the output mode of the collimator. Fig.~\ref{fig:designs}(right), we show the optimized concentrator, significantly increases the the intensity for $-10\degree$ and $10\degree$, which are the angles from both intervals that are closest to the normal incidence, leading to a $2.6\times$ improvement.

\begin{figure}
    \centering
    \includegraphics[width=\textwidth]{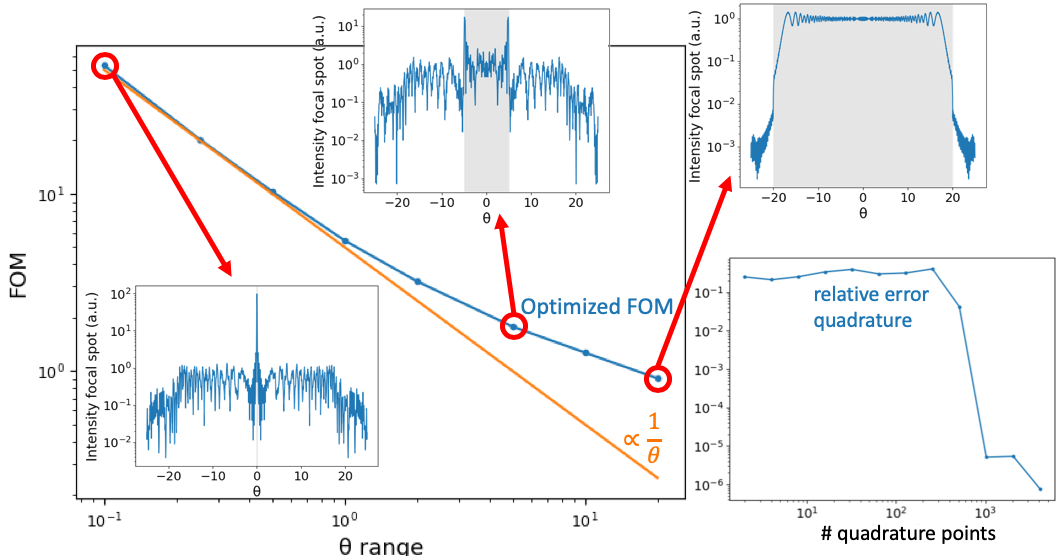} 
    \caption{(left) Optimized average intensity as the range of $\theta$ increases. For $\theta$ up to $1\degree$, the average intensity follow a trend $\propto \frac{1}{\theta}$. (insets) The optimization spontaneously finds multiple strategies depending on the range of the interval. For small range, the design is ``lens-like'' with a maximum intensity at normal incidence. As the range increases, the optimal design chooses the two ends of the interval to be the most intense, for example for the range $[-5\degree, 5\degree]$. For larger interval like $[-20\degree, 20\degree]$, the design tries to performs equally well with low intensity across the interval. (right) Using Gauss--Legendre quadrature, $\sim 10^3$ points are needed to obtain an accurate value of the average. Which corresponds to $\sim 10^3$ Maxwell solves instead of one with our reciprocal method.}
    \label{fig:tradeoffs}
\end{figure}

In Fig~\ref{fig:tradeoffs}(left) we show the optimized average intensity as the range of $\theta$ increases. For $\theta$ up to $1\degree$, the average intensity follow a trend $\propto 1/\theta$---in this small range, the design is ``lens-like'' with a maximum intensity at normal incidence (left inset), and averaging over more angles simply decreases the average linearly. As the angular range increases, however, the design changes.  For moderate ranges, the optimal design chooses the two ends of the interval to be the most intense (middle inset), for example for the range $[-5\degree, 5\degree]$. For even larger ranges like $[-20\degree, 20\degree]$, the design tries to perform equally well with low intensity across the interval (upper-right inset), similar to a min--max optimization~\cite{boyd2004convex}.   In general, it is known that the average intensity must decrease as the angular range is increased, and one theoretical ``brightness bound'' for this tradeoff showed that the upper bound scales inversely with the number of input ``channels''~\cite{zhang2019scattering}.   This channel-inverse scaling is qualitatively similar to the $1/\theta$ scaling that we observe for small angular ranges, but it must have a larger constant coefficient. \update{A single-layer metasurface designed via LPA cannot attain maximal focusing even at a single angle, and typically loses at least 40\% of the power compared to an ideal lens ~\cite{li2022inverse, munley2022inverse, bayati2022inverse}, in part because LPA does not capture the full scattering degrees of freedom in Maxwell’s equations~\cite{chung2020high}, combined with the fact that a single nanopillar layer does not have enough degrees of freedom to eliminate interface reflection. However, with other approximations that capture the effects of bigger unit cells, higher efficiencies ($\approx$70\%) have been demonstrated for single-layer metasurfaces ~\cite{ndao2020octave, phan2019high}.}

\subsection{Brute-force quadrature}

In contrast to our reciprocity approach, the most direct method for modeling incoherent light would be a brute-force average over simulations for many incident-planewave angles.  At the simplest level, this could take the form of a Monte-Carlo method, averaging over many incident waves drawn at random from the given distribution.   Since the desired result \eqref{eq:forwarduncorr} is an integral of planewave-simulations over the angle $\theta$, in which the integrand is a smooth function of $\theta$, however, there are much more effective numerical integration (``quadrature'') techniques.  Even for the most sophisticated high-order Gauss--Legendre quadrature schemes~\cite{trefethen2019approximation}, however, we find that an unreasonable number of simulations are required to attain reasonable accuracy.

In particular, we investigate the required number $N$ of angles~$\theta$ (``quadrature points'') for accurate Gauss--Legendre quadrature of \eqref{eq:forwarduncorr} for the concentrator application above.   For each $N$, we evaluate the relative error (in the mean focal intensity) compared to a high-precision result computed via an adaptive Gauss--Kronrod scheme~\cite{gautschi2004orthogonal} with a low tolerance.  The result is shown in Fig.~\ref{fig:tradeoffs}(right): We find that the relative error is unacceptably large for $\le 512$ quadrature points, and only drops below 5\% for $\ge 1024$ points (after which the error falls exponentially as expected for smooth integrands~\cite{trefethen2019approximation}). We report a similar convergence plot to compute the power in the output mode of a collimator.  That is, a direct integration would require around $1000$ Maxwell simulations (one for each incident quadrature angle $\theta$), whereas our reciprocal approach requires just one. With a single reciprocal simulation, our framework has $4\%$ error compared to the high-precision average. The discrepancy can be explained by the difference of approximation between the forward simulations and the reciprocal simulation, which uses a paraxial approximation.

\section{Concluding remarks}\label{sec:ccl}

This work paves the way for inverse design of optical devices for incoherent illumination, which can impact many real-world applications.
\update{We illustrated our technique by designing single-layer metasurfaces, but the same optimization framework would apply to multilayer metasurfaces, which larger optical volumes would likely lead to more design degrees of freedom and better performance~\cite{zhou2018multilayer, zhou2019multifunctional, mansouree2020multifunctional}.}
Although the examples and formulas in this paper were in 2D for simplicity of illustration, the same approach is directly applicable to 3D, the main difference being that in 3D one needs to account for different polarizations (e.g. three reciprocal simulations for a focal spot, or two for a collimator, if no symmetry is imposed on the metasurface). In 3D, the computational benefit of this technique over brute-force averaging is even more significant, because the naive approach would need to integrate over \emph{two} incident angles (requiring many more quadrature points).
Collimators for incoherent light can be used for white organic light-emitted diodes (WOLED)~\cite{zhou2018light} or chemical bioluminescence-based sensors~\cite{khaoua2021stochastic, roda2016progress}, for example. \update{However, the performance characterization of previous work does not directly compare to our performance characterization as a function of angle.} Concentrators have been of great interest for solar energy~\cite{van2008luminescent}. (For broadband devices such as solar cells, there is still a need for a frequency integral, but this integral can be computed efficiently using a weighted quadrature scheme designed for the solar spectrum~\cite{johnson2019accurate}. One can also apply the same reciprocal technique in the time domain to compute many frequencies simultaneously~\cite{hammond2022high, miller2012photonic}.)   The concentrator and collimator designs in this paper were for a single output spot or a single output mode, but we have developed more general trace-optimization techniques for efficient averages over many spots or modes~\cite{yao2021trace}.

% \section{Data availability} Data underlying the results presented in this paper are available in Dataset 1, Ref. \#.

\section{Acknowledgments}
We thank Prof.~Eli Yablonovitch for helpful discussions on EUV lithography that motivated this work. 

\section{Funding Sources} 
R.P.~was supported by the U.S. Army Research Office through the Institute for Soldier Nanotechnologies (Award No. W911NF-18-2-0048) and by the MIT-IBM Watson AI Laboratory (Challenge No. 2415).  W.Y.~was also supported in part by a grant from the Simons Foundation.

%%%%%%%%%% If using BibTeX:
\bibliography{refs.bib}

\end{document}